\documentclass[a4paper,11pt]{article}
\usepackage{pos}
\usepackage{wrapfig}

\title{Energy loss and equilibration of jets in a QCD plasma}
\author[a]{S. Schlichting}
\author*[a]{I. Soudi
\footnote{\textbf{Acknowledgment:} We thank X.~Du, C.~Greiner, A.~Mazeliauskas, Y.~Mehtar-Tani, G.~D. Moore, B.~Schenke, N.~Schlusser and D.~Teaney for insightful discussions throughout this project.
This work was supported by the Deutsche Forschungsgemeinschaft (DFG, German Research Foundation) – project number 315477589 – TRR 211.
The authors also gratefully acknowledge computing time provided by the Paderborn Center for Parallel Computing (PC$^2$).}}
\emailAdd{isma@physik.uni-bielefeld.de}
\affiliation[a]{Universität Bielefeld,\\
  D-33615 Bielefeld, Germany}

\abstract{We investigate the medium induced fragmentation of jets in a high-temperature QCD plasma. Based on an effective kinetic theory of QCD, we study the non-equilibrium evolution of the jet shower and the chemical equilibration of jet fragments in the medium. By including radiative emissions as well as elastic interactions, our approach extends all the way from the jet energy scale to the temperature of the medium and includes important effects such as the recoil of the medium. We present results for the in-medium fragmentation, including chemical and kinetic equilibration of the soft fragments and discuss implications of our result to jet quenching physics and the problem of thermalization of the quark-gluon plasma in heavy ion collisions.}

\FullConference{%
  HardProbes2020\\
  1-6 June 2020\\
  Austin, Texas}

\begin{document}
\maketitle

\section{Introduction}
The formation of the Quark-Gluon-Plasma (QGP) in the aftermath of heavy-ion collisions has become a great focus of modern QCD studies. High energetic particles, known as jets, provide an important tool to probe the QGP medium. While different approaches are used to study the evolution of jets in the medium, during this work we use an effective kinetic theory, where we include both elastic interaction in small angle approximation and medium induced radiation. Providing us with a description of the full in-medium evolution of jet particles, from high energy domains ($\sim E$) all the way to the thermal bath ($\sim T$). 
\section{Effective kinetic description of in-medium jet evolution}
Based on an effective kinetic description \cite{AMY}, we study the evolution of the energy distribution of jet partons $D_a(x,\tau) \equiv x \frac{d N_a}{dx} $ in a static thermal medium, where $x\equiv \frac{p}{E}$ is the momentum fraction carried by each parton ($a=g,q_f,\bar{q}_f$) and $E$ the total jet energy. Since the hard partons are dilute compared to the soft thermal particles, we will linearize the evolution equation in terms of the jet distribution.
At leading order in coupling constant number conserving $2\leftrightarrow2$ interactions and radiative $1\leftrightarrow 2$ interactions have to be taken into account, leading to the following evolution equation
\begin{eqnarray}
    \partial_{\tau } D_a(x,\tau) &=& C^{1\leftrightarrow 2}_a[\{D_i\}]+C^{2\leftrightarrow 2}_a[\{D_i\}]\;.
\end{eqnarray}
\paragraph{Diffusion approximation}
By introducing a cutoff $\mu$ in the integration of the exchange momentum for the elastic processes, we separate the interactions into large and small angle scatterings 
\begin{eqnarray}
    C^{1\leftrightarrow 2}_a[\{D_i\}] &=& C^{\rm large}_a[\{D_i\}]+C^{\rm small}_a[\{D_i\}]\;.
\end{eqnarray}
Large-angle elastic scatterings exhibit the same parametric dependencies as small angle processes \cite{AMY,Ghiglieri:2015ala}. Therefore,  we will only consider small-angle scatterings in the following and leave the inclusions of large-angle scatterings as a future task.
The small angle scattering can be written in the small angle approximation as a Fokker-Planck equation \cite{Schlichting:2020lef,Blaizot:2015wga,Ghiglieri:2015ala}
\begin{eqnarray}
C_a^{\rm small}[\{f_i\}]=- \nabla_p \mathcal J_a[\{D_i\}] + S_a[\{D_i\}] -\nabla_p\delta \mathcal J_a[\{D_i\}] + \delta S_a[\{D_i\}] \;. \label{eq:Fokker}
\end{eqnarray}
Due to the linearization of the Fokker-Planck equation around the equilibrium distribution, we obtain two types of contributions. The Fokker-Planck operator acting on the jet distribution leads to diffusion, drag $(\nabla_p \mathcal J_a[\{D_i\}])$ and conversion  $(S_a[\{D_i\}])$ of the jet particles. Conversely, the Fokker-Planck operator acting on the equilibrium distribution $(\nabla_p\delta \mathcal J_a[\{D_i\}], \delta S_a[\{D_i\}])$ corresponds to the recoil response of the medium, which describes how the energy lost from the hard sector is deposited into the softer medium particles. 

\paragraph{Collinear radiation}
Hard particles inside the medium undergo multiple soft scattering which induce collinear radiation \cite{AMY}. The infinitely many diagrams generated can be resummed into an effective rate $\frac{d\Gamma^a_{bc}(xE,z)}{dz}$, giving rise to an effective $1\leftrightarrow 2$ collision integral.
The four processes $g\leftrightarrow gg$, $q\leftrightarrow gq$, $\bar{q}\leftrightarrow g\bar{q}$ and $g\leftrightarrow q\bar{q}$ are allowed and it is straightforward to write down their collision integrals. We include both radiation processes and inverse processes (mergings), this later becomes relevant at low scales $(\sim T)$. 
We refer the reader to Ref.~\cite{Schlichting:2020lef} for details of the implementation. 
\section{Energy loss and equilibration of quark and gluon jets}
\begin{wrapfigure}{L}{0.47\textwidth}
    \centering
    \includegraphics[width=0.47\textwidth]{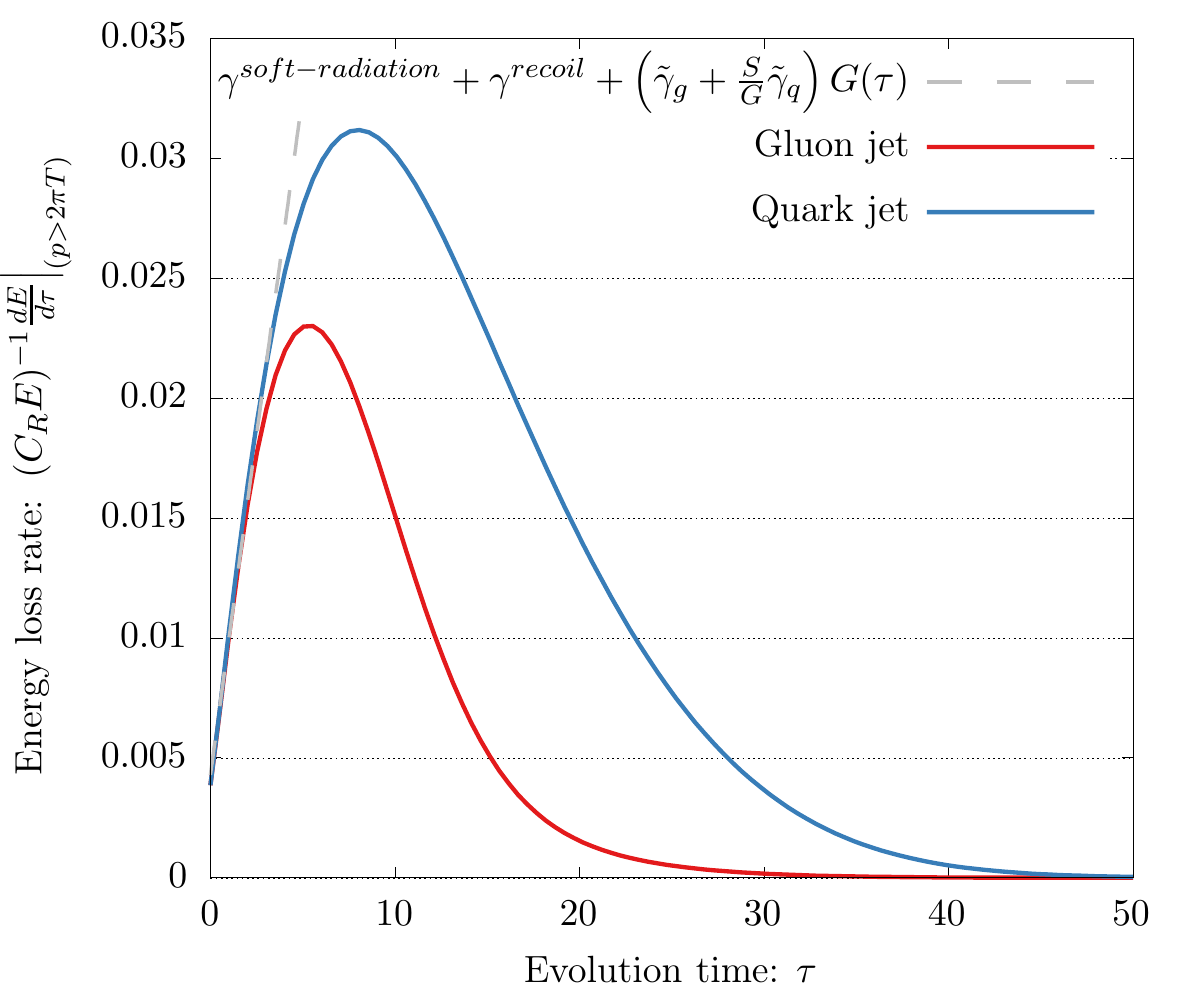}
    \caption{ Differential energy loss rate $dE_{jet}/d\tau$ divided by the corresponding Casimir factor. Dashed curve show the analytical estimate computed in Ref.~\cite{Schlichting:2020lef}. \label{fig:EnergyLoss}}\vspace{-0.5cm}
\end{wrapfigure}
We follow the evolution of an initial jet with energy $E=10^3 T$, where we consider either a gluon or quark carrying all the initial energy ($D_{g/q}^{\rm initial}=\delta(1-x)$). We study the evolution in terms of the dimensionless time variable $\tau \equiv g^4 T\sqrt{\frac{T}{E}} t $, which takes into account the leading jet energy dependence.\\
Since the distribution keeps track of the jet particles even after they completely thermalize, the jet energy is constant throughout the evolution. In order to study energy loss, we define a cutoff scale ($\mu\equiv 2\pi T $) and consider particles with higher energy ($  x > \mu /E $) to be part of the jet. We show in Fig.~(\ref{fig:EnergyLoss}) the evolution of the rate of change by time of the energy carried by these high energy particles for different particle species. 
Based on Fig.~(\ref{fig:EnergyLoss}) we find that the evolution of jet particles inside the medium can be separated into three different stages, characterized by direct energy loss, inverse turbulent cascade, and the eventual approach to equilibrium. We will briefly discuss each regime in the following paragraphs. 

\paragraph{Direct energy loss}
Elastic and inelastic processes both give rise to momentum broadening of the hard jet particles. Although these contributions are small at the early times, they still lead to a considerable deposition of energy in the soft modes $(x\sim T/E)$ due to soft radiation and elastic recoil. One can estimate the deposited energy by computing the energy flux up to the scale $\mu$, leading to a constant energy loss rate \cite{Schlichting:2020lef} shown in Fig.~(\ref{fig:EnergyLoss}). 

\begin{figure}[t!]
    \centering
    \includegraphics[trim={0 0 0 13cm}, clip, width=0.5\textwidth]{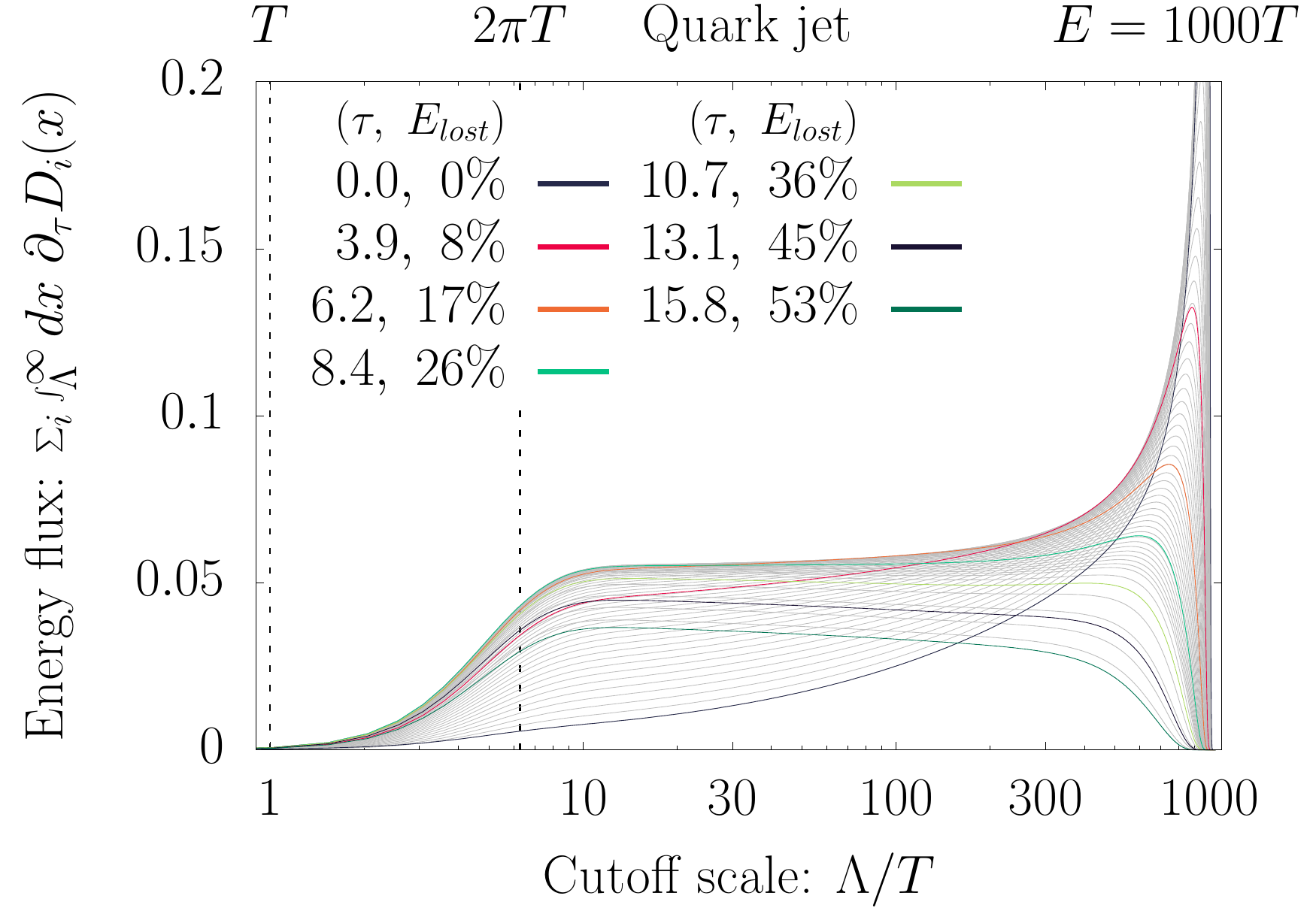}\includegraphics[trim={0 12.5cm 0 0}, clip, width=0.5\textwidth]{figs/EnergyFlux/EnergyFlux-1000.pdf}
    \caption{Evolution of the energy flux in Eq.~\ref{eq:EnergyFlux} for quark (left) and a gluon (right) jets with initial energy $E= 1000T$. Different curves in each panel show the energy flux at different times with gray lines corresponding to intermediate times.}
    \label{fig:EnergyFlux}
\end{figure}
\paragraph{Inverse energy cascade}
The radiative splitting of the initial high energetic patron populate the intermediate energy scales between the jet energy and the medium. These intermediate particles ($T/E \ll x \ll 1$) undergo subsequent splittings with higher splitting rate, leading to an inverse turbulent energy cascade, which dominates the energy loss. In this regime the evolution of jet fragments is governed by inelastic evolution, featuring a stationary solution of the form:
\begin{eqnarray}
    D_g(x) = \frac{G}{\sqrt{x}}\;, \; D_S(x) = \sum_f (D_{q_f} + D_{\bar{q}_f}) = \frac{S}{\sqrt{x}}\;, \; D_V(x) =(D_{q_f} - D_{\bar{q}_f}) =V\sqrt{x}\;,
\end{eqnarray}
which, as demonstrated in earlier work \cite{Blaizot:2015wga,Mehtar-Tani:2018zba}, corresponds to the Kolmogorov-Zhakarov (KZ) spectrum of weak-wave turbulence, and is associated with the stationary transport of energy and valence charge towards lower energies. This spectral shape of the KZ spectrum stems from the characteristic energy fraction $x$-dependence of the splitting rates $\Gamma(xE,z)\sim\sqrt{\frac{\hat{\bar{q}}}{xE}}$ due to the LPM effect. In order to showcase the stationary energy flux, we evaluate the energy flux from high energies up to a cutoff scale $\Lambda$, obtaining
    $\frac{dE}{d\tau}(\Lambda) = \sum_i \int_{\Lambda/E}^\infty dx~ \partial_{\tau} D_i(x)\;.$\label{eq:EnergyFlux}
The energy flux $\frac{dE}{d\tau}(\Lambda)$ is displayed in Fig.~(\ref{fig:EnergyFlux}) for jet energies $E=10^2,10^3T$ at different evolution times. We observe a plateau in the energy flux at intermediate scales of parton energies, where the energy flux is virtually constant during the turbulent regime. Such a constant energy flux indicates that energy is transported from high energy partons $(x\sim 1)$ all the way to the thermal bath $(x\sim T/E)$ without accumulation at intermediate scales. Furthermore, the energy flux showcases the regions where energy is dissipated from the hard particles $(x\sim 1)$ and accumulated at the medium $(x\sim T/E)$.

\begin{figure}
    \centering
    \includegraphics[trim={0 0 12.7cm 0}, clip,width=0.45\textwidth]{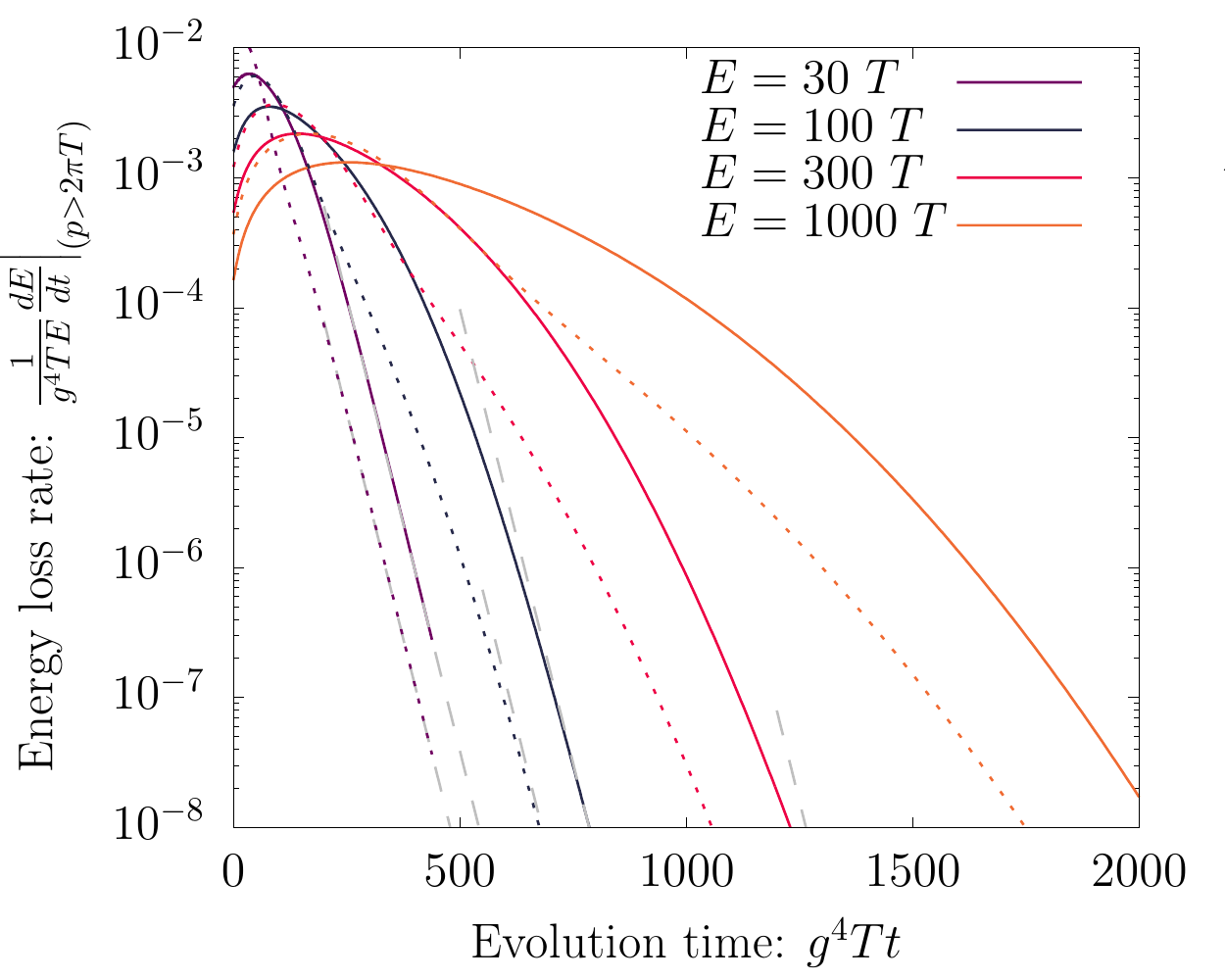}
    \includegraphics[trim={12.7cm 0 0 0 },clip,width=0.45\textwidth]{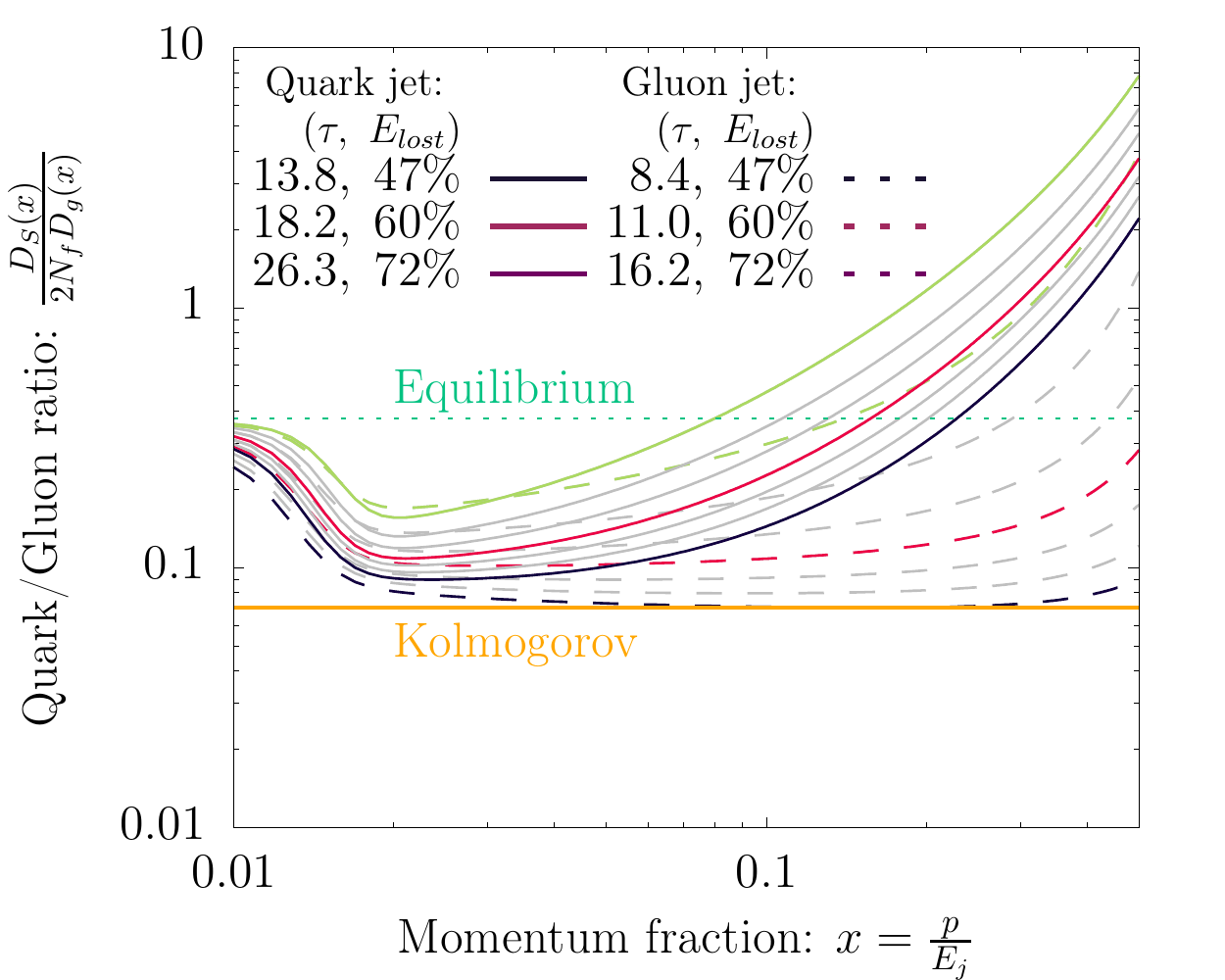}
    \caption{(left) Comparison of energy loss rates for quark (full lines) and a gluon (dashed lines) jets, with different initial jet energies $E=30,100,300,1000$. Dashed lines represent fits to an exponential decay using the first nonzero eigenvalues as the decay constant.
    (right) Ratio of quark to gluon distributions $\frac{D_S(x)}{2N_f D_g(x)}$ for quark (full lines) and gluon (dashed lines) jet at different times $\tau$. Horizontal lines correspond to the equilibrium ratio $D_S(x)/2N_f D_g(x) =\nu_q/\nu_g$ which is approached at small x, and the universal Kolmogorov ratio in \cite{Schlichting:2020lef} which is approached at intermediate values of $T/E \ll x \ll 1$ for a transient period of time.
    \label{fig:ExpDecay}}
    \vspace{-0.2cm}
\end{figure}

\paragraph{Approach to equilibrium}
Eventually all the energy once stored in the hard particles will decay to the thermal bath. At the late stages of the evolution when the jet energy is close to equilibrium, this can be expressed as a small deviation from equilibrium which follows a linear response and relaxes exponentially, with the decay constants computed from the eigenvalues of the linearized collisions operator. In Fig.~(\ref{fig:ExpDecay}) we compare the eigenvalue decay (gray dashed line) to the late decay of the energy loss rate for initial jet energies $E=30,100,300,1000$. We observe that indeed the decay constant matches with the eigenvalues, though, by the time this regime is reached the jets has lost nearly all their energies. 

\paragraph{Effects of jet chemistry}
We observe that the jet chemistry varies as a function of momentum fraction and energy loss as shown in Fig.~(\ref{fig:ExpDecay}). In the infrared region $(x\sim T/E)$, the chemical composition had sufficient time to equilibrate. While in the intermediate scales $(T/E \ll x\ll 1)$ the chemistry follows the universal Kolmogorov ratio \cite{Schlichting:2020lef,Mehtar-Tani:2018zba} at intermediate times.
While at early times the chemistry of the hard sector $x\sim 1$ is dominated by the jet peak, the situation is different at late times.
Because gluons are more efficient at transporting energy from the jet peak to the medium, we observe at late times that even for gluon jets the hard constituents are more likely to be quarks.
\section{Discussion}

Based on an effective kinetic description, we studied the in-medium evolution of jets from the initial energy loss all the way to the equilibration of the jet inside the medium. Due to multiple successive splittings, jet equilibration is dominated by a turbulent cascade which transport energy from the jet sector all the way to the medium. While the in-medium evolution of jets resembles QGP thermalization at early times, due to the large separation of scales between jet energies and the medium, we find that jet quenching is different from near equilibrium physics and therefore allows to probe genuine non-equilibrium features of QCD.

The jet chemistry composition varies throughout the evolution and exhibits interesting effects. Particularly, the medium effectively acts as a chemical filter for gluons, where at late times a higher energetic particle is more likely to be a quark. We expect that these finds will have interesting phenomenological consequence (see e.g. \cite{Schlichting:2020lef,Mehtar-Tani:2018zba}), which can explored further within Monte-Carlo studies of jet-quenching.

\end{document}